\documentclass[10pt,aps,twocolumn,prc,floatfix,preprintnumbers,superscriptaddress,notitlepage,nofootinbib]{revtex4-1}
\usepackage{amssymb, amsmath, bm}

\usepackage[english]{babel}
\usepackage[utf8]{inputenc}
\usepackage{blindtext}
\usepackage{graphicx}
\usepackage{hyperref}
\hypersetup{
    colorlinks,
    linkcolor={red!50!black},
    citecolor={blue!50!black},
    urlcolor={blue!80!black}
}
\usepackage{physics}
\usepackage{xcolor}
\usepackage[normalem]{ulem}
\usepackage{float}
\usepackage{comment}

\usepackage{tikz}

\begin{document}

\title{Restricted Boltzmann Machines Propagators for Auxiliary Field Diffusion Monte Carlo}
\author{Jordan M. R. Fox}
\email{jfox@anl.gov}
\affiliation{Physics Division, Argonne National Laboratory, Argonne, Illinois 60439, USA}
\author{Alessandro Lovato}
\email{lovato@anl.gov}
\affiliation{Physics Division, Argonne National Laboratory, Argonne, Illinois 60439, USA}
\affiliation{Computational Science (CPS) Division, Argonne National Laboratory, Argonne, Illinois 60439, USA}
\affiliation{INFN-TIFPA Trento Institute for Fundamental Physics and Applications, Trento, Italy}

\author{Alessandro Roggero}
\email{a.roggero@unitn.it}
\affiliation{INFN-TIFPA Trento Institute for Fundamental Physics and Applications, Trento, Italy}
\affiliation{Dipartimento di Fisica, University of Trento, via Sommarive 14, I–38123, Povo, Trento, Italy}

\author{Ermal Rrapaj}
\email{ermalrrapaj@berkeley.edu}
\affiliation{Lawrence Berkeley National Laboratory, One Cyclotron RD, Berkeley, CA, 94720, USA}
\affiliation{Department of Physics, University of California, Berkeley, CA 94720, USA}
\affiliation{RIKEN iTHEMS, Wako, Saitama 351-0198, Japan}

\date{\today}

\begin{abstract}
The auxiliary field diffusion Monte Carlo method uses imaginary-time projection techniques to accurately solve the ground-state wave function of atomic nuclei and infinite nuclear matter. In this work, we present a novel representation of the imaginary-time propagator based on restricted Boltzmann machines. We test its accuracy against the routinely employed Hubbard-Stratonovich transformations by evaluating ground-state energies and single-particle densities of selected light nuclei. This analysis paves the way for incorporating more realistic nuclear potentials in the auxiliary field diffusion Monte Carlo method, including isospin-dependent spin-orbit terms and cubic spin-isospin operators, which characterize accurate phenomenological and chiral effective field theory Hamiltonians. 
\end{abstract}

\maketitle

\section{Introduction}
Continuum quantum Monte Carlo methods have a long history of success in solving atomic nuclei using high-resolution Hamiltonians, either phenomenological~\cite{Wiringa:1994wb,Pieper:2001ap} or derived within chiral effective field theory~\cite{Lynn:2015jua,Piarulli:2016vel,Lynn:2017fxg,Piarulli:2017dwd}. In particular, the Green’s function Monte Carlo (GFMC) method~\cite{Carlson:1987zz,Lee:1992,Carlson:2014vla} can solve the nuclear Schr\"odinger equation with percent-level accuracy for both the ground- and low-lying excited state energies of light nuclei. Because the GFMC involves a sum over all spin and isospin states, its computational requirements grow exponentially with the number of particles, presently limiting its applicability to $A\leq 12$ nuclei. The auxiliary field diffusion Monte Carlo (AFDMC)~\cite{Schmidt:1999lik} method has emerged as a powerful alternative to accurately compute nuclei with up to $A\simeq 20$~\cite{Martin:2023dhl,Gnech:2024qru}. To reduce the computational cost from exponential to polynomial in $A$, the spin-isospin degrees of freedom are represented as a tensor product of single-particle spinors. Hubbard-Stratonovich (HS) transformations are employed to make the quadratic spin-isospin-dependent terms entering realistic nuclear forces amenable to this representation. The current main limitation of the AFDMC lies in its inability to encompass highly realistic Hamiltonians, except for purely neutron systems. In fact, the standard HS transformation cannot encompass isospin-dependent spin-orbit terms nor cubic spin-isospin terms entering nucleon-nucleon and three-nucleon forces, respectively~\cite{Gandolfi:2020pbj}. 

Generalizations of this approach have been proposed in the past, including using the HS transformation recursively, at the price of lower efficiency, to accommodate isospin-dependent spin-orbit terms~\cite{zhang2014spin} and making the auxiliary fields self-interacting to allow for the description of many-body interactions~\cite{Korber:2017emn}. It is, of course, possible to linearize the interactions using discrete auxiliary variables instead. A notable example is the Hirsch transform widely used for the Hubbard model~\cite{Hirsch:1983} or the discrete transformations commonly employed in lattice effective field theory simulations~\cite{LEE2009117}. In this work, we build upon the formalism developed in Ref.~\cite{Rrapaj:2020txq}, in which Restricted Boltzmann Machines (RBM) have been shown to generalize these discrete auxiliary field transformations, reproducing the previous construction as special cases while allowing for linearization of arbitrary many-body interactions in the imaginary-time propagator. This RBM representation has a number of advantages over standard HS, most importantly the flexibility to include higher-order operators. While the explicit examples presented in Ref.~\cite{Rrapaj:2020txq} were focused on simple spin/isospin operators, here we generalize the RBM representations to encompass interactions appearing in high-resolution nuclear Hamiltonians, which couple non-trivially the spatial coordinates and spin-isospin degrees of freedom. 

As a proof of principle, we consider the phenomenological Argonne $v_6^\prime$ (AV6P) nucleon-nucleon interaction~\cite{Wiringa:2002ja}, which is highly non-perturbative and entails strong tensor components. We compute the ground-state properties of $A \leq 4$ nuclei, including energies, radii, and single-particle densities. To gauge the accuracy of the RBM representation of the imaginary-time propagator, we benchmark our results against the conventional HS transformation and the highly accurate hyper-spherical harmonics (HH) method~\cite{Kievsky:2008es,Marcucci:2019hml} using the same Hamiltonian as input. The results presented in this work were obtained by implementing the RBM propagator into the massively parallel AFDMC production code written in FORTRAN 2003. To ensure the reproducibility of our results and make the algorithmic developments readily accessible to the broader nuclear and condensed-matter physics community, we developed a Python library suitable for benchmarking our results in the two-nucleon sector. This library is publicly available for download as a GitHub repository~\cite{Fox_spinbox_2024} and is discussed briefly in Appendix~\ref{app:spinbox}.

This paper is organized as follows. In Section~\ref{sec:methods}, we discuss the Hamiltonian and the AFDMC method, with particular emphasis on the imaginary-time propagator. Section~\ref{sec:results} is devoted to benchmarking the RBM propagator against conventional AFDMC results and the HH method. Finally, in Section~\ref{sec:conclusions}, we draw our conclusions and provide perspectives for future applications of our work. 

\section{Methods}
\label{sec:methods}

\subsection{Nuclear Hamiltonian}
In this work, we employ the non-relativistic nuclear Hamiltonian
\begin{align}
H = \sum_i \frac{{\bf p}_i^2}{2m} + \sum_{i<j}^{A} v_{ij} \ ,
\label{H:A}
\end{align}
where ${\bf p}_i$ and $m$ denote the momentum of the $i$-th nucleon and its mass, and $v_{ij}$ is the nucleon-nucleon (NN) potential. In this work, we employ the Argonne $v_6^\prime$ (AV6P) nucleon-nucleon interaction~\cite{Wiringa:2002ja}, which is a projection of the highly-realistic Argonne $v_{18}$ NN potential~\cite{Wiringa:1994wb} onto the first six spin-isospin operators
\begin{equation}
v_{ij} = \sum_{p=1}^6 v_p(r_{ij})O^p_{ij}\,,
\label{eq:AV6P}
\end{equation}
with
\begin{equation}
O^{p=1-6}_{ij}= (1, \sigma_{ij},S_{ij})\otimes (1,\tau_{ij} ) \, .
\label{eq:oper}
\end{equation}
In the above equation we introduced $\sigma_{ij}={\bm \sigma_i}\cdot{\bm \sigma_j}$ and $\tau_{ij}={\bm \tau_i}\cdot{\bm \tau_j}$ with ${\bm \sigma}_i$ and ${\bm \tau}_i$ being the Pauli matrices acting in the spin and isospin space. The tensor operator is given by
\begin{equation}
S_{ij}= \frac{3}{r^2_{ij}}({\bm \sigma}_i\cdot {\bf r}_{ij})({\bm \sigma}_j \cdot{\bf r}_{ij})- \sigma_{ij}\, ,
\end{equation}
In addition, we assume the electromagnetic component of the NN potential to only include the Coulomb force between finite-size protons~\cite{Wiringa:1994wb}.  

The spin/isospin independent term of the NN potential is simply given by the first operator in the sum appearing in Eq.~\eqref{eq:AV6P}: $V_{\rm SI}(R) = \sum_{i<j}^A v_1(r_{ij})$. On the other hand, the spin/isospin-dependent contributions can be expressed as
\begin{align}
V_{\rm SD}(R) &= \sum_{i<j}^A \sum_{p=2}^6 v_p(r_{ij})O^p_{ij} \nonumber \\
&= \sum_{i<j}^A\Big[A_{ij}^{(\tau)} \tau_{ij} +
A_{i\alpha j\beta}^{(\sigma)} \sigma_i^\alpha \sigma_j^\beta + 
A_{i\alpha j\beta}^{(\sigma \tau)} \sigma_i^\alpha \sigma_j^\beta \tau_{ij}\Big]
\end{align}
where a sum over the repeated indexes $\alpha$ and $\beta$ is understood. The real symmetric matrices $A^{(\tau)}$, $A^{(\sigma)}$, and $A^{(\sigma \tau)}$ have dimensions $(A \times A)$, $(3A \times 3A)$, and $(3A \times 3A)$, respectively. 

\subsection{Imaginary-time propagator}
The AFDMC method leverages an imaginary-time propagation to project out the ground-state component of a given Hamiltonian starting from a suitable variational state~\cite{Schmidt:1999lik}
\begin{equation}
|\Psi_0\rangle = \lim_{\tau \to \infty} |\Psi(\tau)\rangle = \lim_{\tau \to \infty} e^{-(H-E_V)\tau} |\Psi_V\rangle\, .
\end{equation}
In the above equation, $E_V$ is a normalization constant, which is chosen to be close to the true ground-state energy $E_0$. The variational ansatz employed in this work is the same as in Ref.~\cite{Gnech:2024qru} and can be schematically written as $|\Psi_V\rangle = \hat{F} |\Phi\rangle$. The mean-field part of the wave function is a Slater determinant of single-particle orbitals, which are parametrized in terms of cubic splines~\cite{Contessi:2017rww}. 
The correlation operator includes a pair-wise product of two- and three-body spin/isospin-dependent Jastrow correlations. To keep a polynomial cost in the number of nucleons, as in Ref.~\cite{Gandolfi:2014ewa}, only linear terms in the spin/isospin correlation operators are kept. The optimal values of the variational parameters are found employing the linear optimization method~\cite{Toulouse:2007,Contessi:2017rww}. 

The imaginary-time propagator $e^{-(H-E_V)\tau}$ is broken down in $N$ small time steps $\delta \tau$, with $\tau= N \delta\tau$. At each step, the spatial and spin/isospin coordinates $R=\{\mathbf{r}_1,\dots,\mathbf{r}_A\}$ and $S=\{s_1,\dots, s_A\}$ are sampled from the previous ones according to the short-time propagator  
\begin{equation}
G(R^\prime, S^\prime, R, S, \delta \tau) = \langle R^\prime S^\prime| e^{-(H-E_V)\delta\tau}|  R S\rangle\,.
\end{equation}

Employing the Suzuki-Trotter decomposition, one can separate the kinetic, $T$ and potential $V$, components of the Hamiltonian as
\begin{equation}
e^{-(T + V)\delta\tau } = e^{-V\delta\tau / 2 } e^{-T\delta\tau } e^{-V\delta\tau / 2 } + \mathcal{O}(\delta\tau^3)\,.
\end{equation}

The kinetic energy gives rise to the free-particle propagator that can be expressed as a simple Gaussian in configuration space, and it is diagonal in the spin/isospin space
\begin{equation}
\begin{split}
&\langle R^\prime S^\prime | e^{-T \delta\tau } |R S\rangle \\
& = \Big(\frac{m}{2\pi\delta\tau}\Big)^{3A/2} e^{-\frac{m(R-R^\prime)^2}{2\delta\tau}} \delta(S-S^\prime)\,.
\end{split}
\end{equation}
On the other hand, for potentials that are local in coordinate space, the corresponding propagator is diagonal in the coordinate space but not in the spin/isospin one. 
\begin{equation}
\begin{split}
& \langle R^\prime S^\prime | e^{-V \delta\tau / 2 } |R S\rangle \\
& = e^{-V_{\rm SI} (R) \delta\tau / 2 } \langle S^\prime | e^{-V_{\rm SD} (R) \delta\tau / 2 } | S\rangle\, \delta(R-R^\prime)
\end{split}
\label{eq:prop_v}
\end{equation}
The AFDMC spin-isospin basis is given by the tensor product of single-nucleon spinors
\begin{equation}
|S\rangle = |s_1\rangle \otimes |s_2\rangle \otimes \cdots \otimes |s_A\rangle\,.
\end{equation}
Since $V_{\rm SD}$ contains terms that are quadratic in spin/isospin Pauli matrices, the exponential $e^{-V_{\rm SD} (R) \delta\tau / 2 }$ connects $|S\rangle$ to an exponentially-large number of single particle spinors $|S^\prime\rangle$. To preserve the tensor-product representation above, the quadratic spin-isospin term entering the NN potential is linearized, introducing a set of fluctuating auxiliary fields.

\subsubsection{Conventional Hubbard-Stratonovich transformation}
Within the conventional HS approach, the real and symmetric matrices defining $V_{\rm SD}$ are first diagonalized as
\begin{align}
\sum_{j} A_{ij}^{(\tau)} \Psi_{n,j}^{(\tau)} &= \lambda_n ^{(\tau)} \Psi_{n,i}^{(\tau)}\,,\nonumber\\
\sum_{j\beta} A_{i\alpha j\beta}^{(\sigma)} \Psi_{n,j\beta}^{(\sigma)} &= \lambda_n ^{(\sigma)} \Psi_{n,i\alpha}^{(\sigma\tau)}\,,\nonumber\\
\sum_{j\beta} A_{i\alpha j\beta}^{(\sigma\tau)} \Psi_{n,j\beta}^{(\sigma\tau)} &= \lambda_n ^{(\sigma\tau)} \Psi_{n,i\alpha}^{(\sigma\tau)}\, .
\end{align}
Then, the following operators are defined in terms of their eigenvectors
\begin{align}
\mathcal{O}_{n\alpha}^{(\tau)} &=  \sum_{j} \tau_j^\alpha \Psi_{n,j}^{(\tau)} \, \nonumber\\
\mathcal{O}_n^{(\sigma)} &=  \sum_{j\beta} \sigma_j^\beta \Psi_{n,j\beta}^{(\sigma)} \, \nonumber\\
\mathcal{O}_{n\alpha}^{(\sigma\tau)} &=  \sum_{j\beta} \tau_j^\alpha \sigma_j^\beta \Psi_{n,j\beta}^{(\sigma\tau)} \,,
\end{align}
so that
\begin{align}
V_{\rm SD}(R) &= \frac{1}{2}\sum_{\alpha=1}^3\sum_{n=1}^{A} \lambda_n^{(\tau)}\left(\mathcal{O}_{n\alpha}^{(\tau)}\right)^2 +  \frac{1}{2} \sum_{n=1}^{3A} \lambda_n^{(\sigma)} \left(\mathcal{O}_n^{(\sigma)}\right)^2 \nonumber\\
&+\frac{1}{2}\sum_{\alpha=1}^3\sum_{n=1}^{3A} \lambda_n^{(\sigma \tau)}\left(\mathcal{O}_{n\alpha}^{(\sigma\tau)}\right)^2\,.
\label{eq:V_SD}
\end{align}
The quadratic operators are then linearized, leveraging the HS transformation
\begin{equation}
e^{-\frac{1}{2} \lambda \mathcal{O}^2} = \frac{1}{\sqrt{2 \pi}}\int_{-\infty}^\infty dx\, e^{-\frac{x^2}{2} + \sqrt{-\lambda} x \mathcal{O}}
\label{eq:hs}
\end{equation}
where $x$ is usually called the {\it auxiliary field}. 
Hence, the spin/isospin-dependent part of the potential-energy propagator of Eq.~\eqref{eq:prop_v} can be expressed as
\begin{equation}
\begin{split}
\langle S^\prime | e^{-V_{\rm SD} (R) \delta\tau / 2 } | S\rangle &= \prod_{n=1}^{N} \frac{1}{\sqrt{2 \pi}} \int_{-\infty}^\infty dx_n e^{-\frac{x_n^2}{2}} \\
& \times \langle S^\prime | e^{\sqrt{- \frac{\delta\tau}{2} \lambda}\, x_n \mathcal{O}_n} |S\rangle\,.
\end{split}
\end{equation}
where $N=15A$ and we have neglected terms of order $\delta\tau^2$ coming from $[\mathcal{O}_m, \mathcal{O}_n]$. 
The above integral is approximated by a finite sum of samples of $x_n$ drawn from a standard normal distribution. To exactly cancel terms that are linear in the auxiliary fields and in $\sqrt{\delta\tau}$, for a given set of sampled $X = \{x_1, \dots, x_N\}$ we also consider the sample obtained flipping the sign of the auxiliary fields $-X = \{-x_1, \dots, -x_N\}$. An importance-sampled heat-bath procedure is then employed to keep only one sample in the imaginary-time propagation~\cite{Piarulli:2019pfq, Gandolfi:2020pbj}. 

In keeping a single Gaussian sample $X$ from the integral above, one obtains a rotated spin state $|S^\prime(X)\rangle$, which is still a tensor product of single-particle spinors. The entanglement introduced by the quadratic spin/isospin terms in the NN potential is recovered only when expectation values are stochastically estimated by averaging over the Monte Carlo samples.

\subsubsection{Restricted Boltzmann machine transformation}

Notable progress has been made in recent years exploring the connection between Boltzmann machines and strongly correlated spin systems \cite{2021arXiv210315917B, PhysRevB.100.064304, Decelle_2021, Decelle_2024, YEVICK2021107518, PhysRevB.96.205152, SciPostPhys.12.5.166}.
In machine-learning contexts, the RBM is employed as an energy-based neural network model to learn a probability distribution, but this is in many ways identical to the traditional statistical treatment of an Ising model \cite{PhysRevLett.35.1792}. 
The simplest RBM propagator may even be interpreted as a promotion of the scalar auxiliary field of HS to a quantum spin-up/spin-down variable.

For a general RBM network with $n$ visible variables $\bm{v}\in\mathbb{R}^n$ and $m$ hidden discrete-valued variables $\bm{h}\in \Omega = \{0,1,2,\dots,\mathcal{K}-1\}^m$, the partition function $Z_\text{RBM}$ is defined as
\begin{equation}
    Z_\text{RBM}= \mathcal{N} \sum_{\bm{h}\in\Omega} \exp \{ - F_\text{RBM}(\bm{h})\}
\end{equation}
where $\mathcal{N}$ is a normalization constant and
\begin{equation}\label{eq:f_rbm}
    F_\text{RBM}(\bm{h}) = \sum_{i=1}^n \sum_{j=1}^m b_i v_i + c_j h_j + v_i W_{ij} h_j
\end{equation} 
is the free energy of the system, which depends on the network weights $\bm{b},\bm{c}$ (vectors), and $W$ (a matrix).

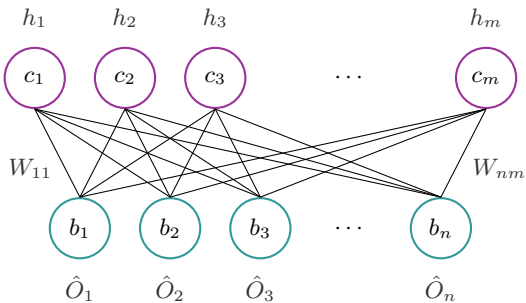
\begin{figure}
    \begin{tikzpicture}
    \tikzset{cir0/.style={circle,draw=teal!80,thick,minimum size=0.8cm},y=0.6cm,font=\sffamily}
    \tikzset{cir1/.style={circle,draw=violet!80,thick,minimum size=0.8cm},y=0.6cm,font=\sffamily}
    
    \begin{scope}[rotate=90]
    \node[cir0] (b1) at (0,0) {$b_1$};
    \node[cir0] (b2) at (0,-1*2) {$b_2$};
    \node[cir0] (b3) at (0,-2*2) {$b_3$};
    \node[cir0] (b4) at (0,-4*2) {$b_n$};
    \node[cir1] (c1) at (2,1) {$c_1$};
    \node[cir1] (c2) at (2,-1*2+1) {$c_2$};
    \node[cir1] (c3) at (2,-2*2+1) {$c_3$};
    \node[cir1] (c4) at (2,-4*2-1) {$c_m$};
    
    \draw (-0.8, 0) node[black!80] {$\hat{O}_1$};
    \draw (-0.8, -1*2) node[black!80] {$\hat{O}_2$};
    \draw (-0.8, -2*2) node[black!80] {$\hat{O}_3$};
    \draw (-0.8, -4*2) node[black!80] {$\hat{O}_n$};
    \draw (2.8, 1) node[black!80] {$h_1$};
    \draw (2.8, -1*2+1) node[black!80] {$h_2$};
    \draw (2.8, -2*2+1) node[black!80] {$h_3$};
    \draw (2.8, -4*2-1) node[black!80] {$h_m$};
    \draw (0.8, -4*2-1.3) node[black!80] {$W_{nm}$};
    \draw (0.8, 1.1) node[black!80] {$W_{11}$};
    
    \foreach \i in {0,2} {
        \draw (\i,-3*2) node {$\cdots$};
    }
    
    \foreach \cnto in {1,2,3,4} {
        \foreach \cntt in {1,2,3,4} {
            \draw [-] (b\cnto.north)--(c\cntt.south);
        }
    }    
    \end{scope}
    \end{tikzpicture}    
    \caption{Graph representation of single-layer quantum RBM. The $\hat{O}_i$ are one-body operators, $h_j$ are discrete-valued hidden variables. The parameters $\bm{b},\bm{c}$, and $W$ can be determined analytically such that the free energy of the network effectively replaces physical forces. }
    \label{fig:rbm_graph}
\end{figure}

For a quantum RBM, displayed in Fig.~\ref{fig:rbm_graph}, we promote the visible variables $\bm{v}$ to involutory operators, $\hat{O}^2=1$, and the resulting free energy is given by~\cite{Rrapaj:2020txq}
\begin{equation}\label{eq:f_rbm_sigma}
    \hat{F}_\text{RBM}(\bm{h}) = \sum_{i=1}^n \sum_{j=1}^m b_i \hat{O}_i + c_j h_j + \hat{O}_i W_{ij} h_j
\end{equation}

For this application to nuclear AFDMC, we identify $\hat{O}_i$ with $\sigma_i^\alpha$, $\tau_i^\alpha$, and $\sigma_i^\alpha \tau_j^\beta$. We choose the simplest nontrivial RBM, allotting a single discrete auxiliary variable $h \in \{0,1\}$ ($m=1$, $\mathcal{K}=2$) to each term entering $V_{\rm SD}(R)$ of Eq.~\eqref{eq:V_SD}. 

To be definite, let us consider the interaction modeled by the $A_{i\alpha j\beta}^{(\sigma)}$ matrix. The quantum RBM allows us to define the mapping
\begin{equation}\label{eq:matching}
    e^{-\frac{\delta\tau}{2} A_{i\alpha j\beta}^{(\sigma)} \sigma_i^\alpha \sigma_j^\beta} = \mathcal{N} \sum_{h=0,1} e^{ -h (C + W_1 \sigma_i^\alpha + W_2 \sigma_j^\beta)}\,.
\end{equation}
Here, we have neglected the bias ($\bm{b}=0$) because it corresponds to a one-body interaction.
In general, the RBM parameters $\mathcal{N}$, $C$, $W_1$, and $W_2$ could be found using a numerical optimization algorithm for training artificial neural network models. However, in this case, they can also be determined analytically in terms of the strength factor $A_{i\alpha j\beta}^{(\sigma)}$, as discussed in detail in Appendix \ref{app:derivation}. The final result reads 
\begin{equation}
    e^{- z \sigma_i^\alpha \sigma_j^\beta} = \frac{e^{- |z| }}{2} \sum_{h=0,1} e^{ \cosh^{-1} \left( e^{2|z|} \right) ( h-1/2 ) (\sigma_i^\alpha - \frac{z}{|z|} \sigma_j^\beta)  }\,,
\label{eq:g_rbm_2b}
\end{equation}
where, for brevity, we introduced $z = \frac{\delta\tau}{2} A_{i\alpha j\beta}^{(\sigma)}$
The two-body RBM propagator in Eq.~\eqref{eq:g_rbm_2b} works similarly for two-body operators associated with $A_{ij}^{(\tau)}$ and $A_{i\alpha j\beta}^{(\sigma\tau)}$. 
Note that, treating the $A_{i\alpha j\beta}^{(\sigma)}$ term within the RBM propagator requires sampling $9 \times A(A-1)/2$ discrete auxiliary fields. This number has to be confronted with the $3A$ Gaussian auxiliary fields that are needed when $A_{i\alpha j\beta}^{(\sigma)}$ is diagonalized, as in the conventional HS transformation. A similar procedure can be followed for including the $A_{ij}^{(\tau)}$ and $A_{i\alpha j\beta}^{(\sigma\tau)}$ matrices in the RBM propagator, which requires additional $A(A-1)/2$ and  $9 \times A(A-1)/2$ auxiliary fields, respectively.

The availability of analytical expressions of the RBM parameters in terms of $A_{i\alpha j\beta}^{(\sigma)}$ is critical for the practical application of RBM propagators to the AFDMC since the matrix $A_{i\alpha j\beta}^{(\sigma)}$ depends upon the $3A$ spatial coordinates of the nucleons. Thus, finding analytical solutions for the RBM parameters allows us to efficiently compute the exact RBM representation of the imaginary-time propagator associated with $V_{\rm SD}(R)$. 

\section{Results}
\label{sec:results}
To gauge the accuracy of the RBM imaginary-time propagator, we first compute the ground-state energies of $^3$H, $^3$He, and $^4$He using as input the AV6P Hamiltonian and the Coulomb repulsion among finite-size protons. We benchmark these energies against AFDMC calculations based on the HS transformations and the highly accurate hyperspherical harmonics (HH) method~\cite{Kievsky:2008es,Marcucci:2019hml}.

The RBM propagator slightly under-binds $^3$H, $^3$He, and $^4$He compared to HH --- the differences, however, always remain within $2\%$ of the HH value. On the other hand, the conventional HS propagator slightly over-binds these nuclei, especially  $^4$He. 
It must be noted that the AFDMC results corresponding to both the RBM and HH propagators are obtained within the so-called ``constrained-path'' approximation, which is employed to control the fermion-sign problem~\cite{Zhang:1996us,Zhang:2003zzk,Wiringa:2000gb}. 
It has been recently shown that performing unconstrained propagations brings AFDMC energies in much better agreement with the HH ones, at least when the conventional HS propagator is used~\cite{Gnech:2024qru}. 
We expect a similar behavior for the RBM propagator, although numerical confirmation of this hypothesis requires using significant computational resources on the order of one million CPU hours, and it goes beyond the scope of the present work. 

\begin{table}[!htb]
\renewcommand{\arraystretch}{1.25} 
\centering
\begin{tabular}{ c | c c c }
\hline
\hline
       & RBM & HS & HH  \\
\hline
$^2$H  & -2.28(1)  & -2.27(1)  & -2.24(1) \\
$^3$H  & -7.85(2)  & -7.97(2)  & -7.91(1)  \\
$^3$He & -7.12(1)  & -7.28(1)  & -7.26(1)  \\
$^4$He & -26.05(5) & -26.37(6) & -26.13(1) \\
\hline
\end{tabular}
\caption{Ground-state energies in MeV of $^3$H, $^3$He, and $^4$He nuclei obtained with the AV6P NN potential and Coulomb repulsion between finite-size protons. The AFDMC method based on the RBM imaginary-time propagator is compared with the conventional HS propagator, as well as the highly accurate HH few-body technique.~\label{tab:energies_light}}
\end{table}

To further benchmark the RBM propagator, we present results for the point-nucleon density, which is defined as
\begin{align}
	\rho_{N}(r) &=\frac{1}{4\pi r^2}\langle \Psi_0 \big|\sum_i \mathcal \delta(r-|\mathbf{r}_i|)|\Psi_0 \rangle\,.
	\label{eq:rho_N}
\end{align}
Center of mass contaminations are automatically removed by $\mathbf{r}_i \to \mathbf{r}_i -\mathbf{R}_{\rm CM}$, where $\mathbf{R}_{\rm CM} = \sum_{i}\mathbf{r}_i/A$ is the center of mass coordinate of the nucleus~\cite{Massella:2018xdj}. Since this quantity does not commute with the Hamiltonian, its evaluation requires using perturbation theory as
\begin{equation}
\frac{\langle\Psi(\tau) | O | \Psi(\tau)\rangle}{\langle\Psi(\tau) | \Psi(\tau)\rangle} \simeq 
2 \frac{\langle\Psi_V | O | \Psi(\tau)\rangle}{\langle\Psi_V | \Psi(\tau)\rangle} - \frac{\langle\Psi_V | O | \Psi_V\rangle}{\langle\Psi_V | \Psi_V\rangle}\, .
\label{eq:pc}
\end{equation}
Note, however, that the variational state is the same for the RBM and HS propagators.

\begin{figure}[!htb]
    \centering
    \includegraphics[width=\columnwidth]{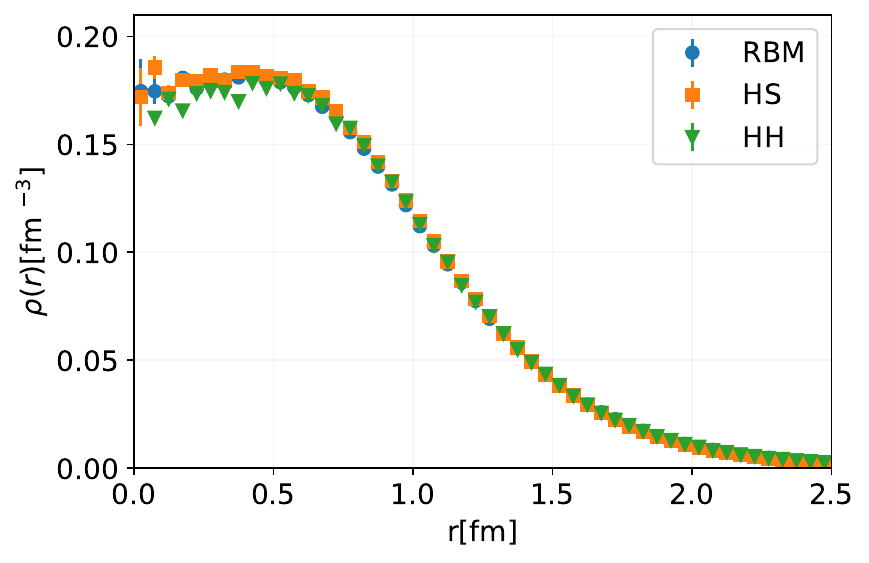}
    \caption{Point nucleon density of $^4$He as obtained with the AFDMC using the RBM and HS propagator for the AV6P Hamiltonian and the HH few-body technique.}
    \label{fig:he4_density}
\end{figure}

Figure~\ref{fig:he4_density} displays the point-nucleon density of $^4$He obtained using the RBM and HS propagators compared with the HH method. We can observe an excellent agreement between them; the density distributions are completely consistent, within errors, over the entire space. The point radii are also compatible, as we get $\sqrt{\langle r_{\rm pt} \rangle} = 1.46(1)$ fm and $\sqrt{\langle r_{\rm pt} \rangle} = 1.44(1)$ fm for the RBM and HS propagators, respectively. 

\section{Conclusions}
\label{sec:conclusions}
In this work, we introduced a RBM representation of the AFDMC imaginary-time propagators, which is amenable to nuclear forces with a strong tensor component. We demonstrated the accuracy of this approach for ground-state energies of  $^3$H, $^3$He, and $^4$He nuclei with the conventional HS propagator routinely used in AFDMC calculations and the highly accurate HH method. In addition to ground-state energies, single nucleon densities also show excellent agreements with those obtained with the conventional HS propagator and the HH technique.  

Although not essential for the AV6P potential employed in this work, employing the RBM propagator will be critical for overcoming a long-standing limitation of the AFDMC, namely its inability to encompass isospin-dependent spin-orbit terms. The latter appear in both highly-accurate phenomenological NN interactions like the Argonne v$_{18}$~\cite{Wiringa:1994wb} and in chiral effective field theory potentials at N3LO~\cite{Piarulli:2014bda,Somasundaram:2023sup}. Moreover, the RBM propagator will allow us to incorporate cubic spin/isospin operators that enter both phenomenological and chiral effective field theory three-nucleon forces~\cite{Pudliner:1995wk,Pieper:2008rui,Lynn:2015jua,Piarulli:2017dwd}. So far, both isospin-dependent spin-orbit terms and cubic spin/isospin operators are included in an approximate way~\cite{Gnech:2024qru}, preventing the applicability of this method to high-precision study of medium-mass nuclei and isospin-symmetric nuclear matter. Reaching these is important to carry out rigorous tests of nuclear Hamiltonians, including phenomenological ones and those derived within chiral effective field theories.  
 
Finally, our work paves the way to using artificial neural networks to efficiently represent the imaginary time propagator entering nuclear diffusion Monte Carlo methods, so far successfully applied to spin systems~\cite{Carleo:2018iag}. In future work, we plan on employing artificial neural networks to represent the AFDMC imaginary-time propagator, which entails both spatial and spin-isospin degrees of freedom for short imaginary times. We foresee a generalization of what has been done with the Green's function Monte Carlo method using a basis expansion~\cite{Pudliner:1995wk}, but not limited to the two-body sector only. Our goal is to be able to use much larger imaginary-time steps, thereby reducing the computational cost of the calculation and access quantities that are notoriously difficult to evaluate accurately, such as electroweak response functions~\cite{Lovato:2016gkq,Lovato:2020kba,Gnech:2024qru}.

We also provide a Python package \texttt{spinbox} for transparent analyses of AFDMC and GFMC internal processes. The tools provided allow for high-level exploration of imaginary-time propagation using HS and RBM implementations, statistical analyses, and more. 
See Appendix \ref{app:spinbox} for details.
Those interested in contributing are encouraged to contact JF.

\section*{Acknowledgments}
We thank Alex Gnech for providing us with the results of the hyper-spherical harmonics method. 
The present research is supported by the U.S. Department of Energy, Office of Science, Office of Nuclear Physics, under contracts DE-AC02-06CH11357 (J.~F., and A.~L.), by the DOE Early Career Award program (A.~L.), and by the SciDAC-NUCLEI (J.~F., and A.~L.) and SciDAC-NeuCol (J.~F., and A.~L.) projects.

\appendix

\section{Derivation of RBM parameters}\label{app:derivation}

Here, we provide the analytic determination of the RBM parameters for the NN propagator; the reader is directed to appendix A of Ref.~\cite{Rrapaj:2020txq} for more. 
For the operator $z \hat{\sigma}_1 \hat{\sigma}_2$, of direct interest for the AFDMC method, we begin with setting the physical propagator equal to the RBM partition function, albeit with effective lower-order terms $z_1 \hat{\sigma}_1, z_2 \hat{\sigma}_2 $ included for the matching procedure
\begin{equation}\label{eq:matching_app}
    e^{-(z \hat{\sigma}_1 \hat{\sigma}_2 + z_1 \hat{\sigma}_1 + z_2 \hat{\sigma}_2 ) } = \mathcal{N} \sum_{h=0,1} e^{ -h (C + W_1 \hat{\sigma}_1 + W_2 \hat{\sigma}_2)}\,.
\end{equation}

Since each spin state can be represented as a linear combination of Pauli operator eigenstates, we can secure a general solution by considering all possible combinations of eigenvalues. Each $\sigma$ has eigenvalues $\pm1$, so there are 4 possible combinations: $(\sigma_1,\sigma_2) \in \{ (+1,+1), (+1,-1), (-1,+1), (-1,-1) \} $. The matching equation is
\begin{equation}
\begin{split}
    &z \sigma_1 \sigma_2 + z_1 \sigma_1 + z_2 \sigma_2\\
    &= - \ln \left[ \mathcal{N} \left( 1 + e^{-C - W_1 \sigma_1 - W_2 \sigma_2} \right) \right]\, .
\end{split}
\end{equation}

Plugging in the four eigenvalue combinations gives this system of equations.
\begin{equation}
  \begin{aligned}
  z + z_1 + z_2  &= - \ln \left( \mathcal{N} \left( 1 + e^{-C - W_1 - W_2} \right) \right) \\
  -z + z_1 - z_2 &= - \ln \left( \mathcal{N} \left( 1 + e^{-C - W_1 + W_2} \right) \right) \\
  -z - z_1 + z_2 &= - \ln \left( \mathcal{N} \left( 1 + e^{-C + W_1 - W_2} \right) \right) \\
  z - z_1 - z_2  &= - \ln \left( \mathcal{N} \left( 1 + e^{-C + W_1 + W_2} \right) \right)\,.
  \end{aligned}
\end{equation}
Dividing the first three equations by the last one cancels the normalization $\mathcal{N}$, yielding a well-determined linear system for three variables:
\begin{equation}
    \begin{pmatrix}
        0 & 2 & 2 \\
        -2 & 2 & 0 \\
        -2 & 0 & 2 \\
    \end{pmatrix}
    \begin{pmatrix}
        z\\
        z_1\\
        z_2\\
    \end{pmatrix}
    =
    \begin{pmatrix}
        L(+1,+1)\\
        L(+1,-1)\\
        L(-1,+1)\, ,
    \end{pmatrix}
\end{equation}

where

\begin{equation}
\begin{split}
    L(\sigma_1,\sigma_2) =& -\ln\left( 1 + e^{-C-W_1\sigma_1-W_2\sigma_2}\right)\\ &+\ln\left( 1 + e^{-C+W_1+W_2}\right)\, .
\end{split}
\end{equation}

The following solutions to the linear system are determined by the symbolic software {\it Mathematica}~\cite{Mathematica}.
\begin{equation}
  \begin{aligned}
  z &= \frac{1}{4} \ln\left(\frac{
   \left(e^{-C+W_1-W_2}+1\right)\left(e^{-C-W_1+W_2}+1\right)}{\left(e^{-C-W_1-W_2}+1\right)
   \left(e^{-C+W_1+W_2}+1\right)}\right)\\
  z_1 &= \frac{1}{4} \ln\left(\frac{\left(e^{-C+W_1-W_2}+1\right)\left(e^{-C+W_1+W_2}+1\right)}{\left(e^{-C-W_1-W_2}+1\right)\left(e^{-C-W_1+W_2}+1\right)}
\right) \\
  z_2 &= \frac{1}{4} \ln\left(\frac{\left(e^{-C-W_1+W_2}+1\right)\left(e^{-C+W_1+W_2}+1\right)}{\left(e^{-C-W_1-W_2}+1\right)\left(e^{-C+W_1-W_2}+1\right)}
\right) \\
  \end{aligned}  
\end{equation}

Recall that $z_1$ and $z_2$ correspond to extra lower-order operators, meaning this system is still under-determined as solutions to the RBM propagator. We choose $C=0$, $|W_1|=|W_2|=W$, which simplifies the above expressions to
\begin{align}
    z_1 &= \frac{W_1}{2}\nonumber\\
    z_2 &= \frac{W_2}{2}\, .
\end{align}
Based on our choice, there are two possibilities for  $W$: either $W_1=W_2=W$ or $W_1=-W_2=W$. If the former, then
\begin{equation}
    W_{(W_1=W_2)} = \cosh^{-1} \left( e^{-2z} \right)\,,
\end{equation}
and if the latter, then
\begin{equation}
    W_{(W_1=-W_2)} = \cosh^{-1} \left( e^{+2z} \right)\,.
\end{equation}
The value of $\cosh^{-1}$ is real for arguments in $[1,+\infty)$, so if $z<0$ then the relative sign of $W_1,W_2$ is +1 (former case), and if $z>0$ then the relative sign of $W_1,W_2$ is -1 (latter case). 
These can be satisfied simultaneously with
\begin{align}
    z_1 &= \frac{W}{2} \nonumber\\
    z_2 &=  -\frac{z}{|z|} \frac{W}{2}\,,
\end{align}
where $W=\cosh^{-1} \left( e^{2|z|} \right)$. Inserting these solutions into Eq.~\eqref{eq:matching_app} leads to the two-body RBM transformation reported in Eq.~\eqref{eq:g_rbm_2b}.

\section{\texttt{spinbox}} \label{app:spinbox}

\begin{figure}
    \centering
    \includegraphics[width=\columnwidth]{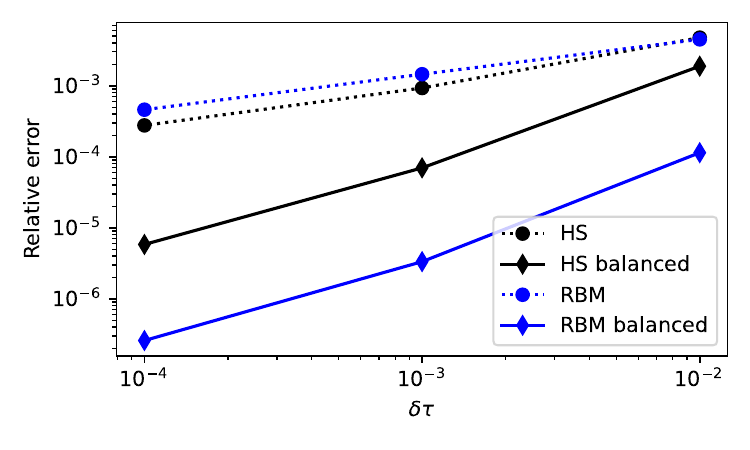}
    \caption{Relative error in an example calculation of two-body propagator amplitude (Eq.~\eqref{eq:amplitude}) as a function of the imaginary-time step $\delta \tau$. The RBM has a significantly smaller error than HS when balanced. }
    \label{fig:error_plot}
\end{figure}

The primary goal of our Python library \texttt{spinbox} \cite{Fox_spinbox_2024} is to provide tools to analyze elements of imaginary time propagation and particularly the internal processes of large AFDMC and GFMC calculations.
For a simulation using tensor-product spin-isospin vectors $|S\rangle$ and $|S'\rangle$, involutory operators $\hat{O}_1,\hat{O}_2$ acting on particles 1 and 2, one needs to calculate amplitudes of the form
\begin{equation}
    \langle S' | e^ {- z \hat{O}_1 \hat{O}_2 } | S \rangle \,,
\end{equation}
with $z \in \mathbb{C}$.
If the full basis of orthogonal spin-isospin tensor-product states is to be used, this calculation can be done by vector-matrix-vector multiplication, using a Pad\'e approximant (for example) to estimate the exponential of the two-body operator matrix. 
However, if one must propagate a single tensor-product state, as is done in AFDMC, then an integral transform like HS or the RBM must be applied.
The set of all tensor-product vectors is not a proper subspace, as it is not closed under addition; consequently, the resulting expression using the HS transformation (for example) must be 
\begin{equation}\label{eq:amplitude}
    \langle S' | e^ {- z \hat{O}_1 \hat{O}_2 } | S \rangle = \lim_{M \rightarrow \infty} \frac{e^z}{M} \sum_{x \sim \mathcal{N}}^M  \langle S' | e^{x \sqrt{-z} (\hat{O}_1 + \hat{O}_2 )} |S \rangle \,,
\end{equation}
where the averaging occurs over scalar values (as opposed to summing vectors).
The equivalent calculation using the RBM follows directly from Eq.~\eqref{eq:g_rbm_2b}.

Figure \ref{fig:error_plot} shows the results of \texttt{spinbox} calculations comparing relative error an example calculation of Eq.~\eqref{eq:amplitude} (a randomly chosen integral for $A=4$ with AV6P using $10^6$ auxiliary field samples). 
``Balanced'' refers to calculations where the average over auxiliary fields is balanced according to the relevant symmetry: $x \rightarrow -x$ for Gaussian, and $h \rightarrow 1-h$ for binary. Our RBM propagator has notably smaller errors than standard HS; over one order of magnitude smaller using balanced samples.

The \texttt{spinbox} package provides the following capabilities.
\begin{itemize}
    \item Numerical representation of spinor samples and operators, including methods for relevant operations. Spin ($\sigma_z = \pm \frac{1}{2}$) or spin-isospin ($\sigma_z = \pm\frac{1}{2}$, $\tau_z = \pm\frac{1}{2}$), and any number of particles. 
    \item Choice of basis space, either one tensor-product spinor (AFDMC) or a linear combination (GFMC). Methods for going between the two bases.     
    \item Hubbard-Stratonovich, RBM, and ``exact'' propagators for NN forces. Currently being extended to NNN.
    \item Built-in AV6P, spin-orbit, and Coulomb forces. Devoted classes and methods for couplings.
    \item Thread-parallel calculation of the average in Eq.~\eqref{eq:amplitude}, and the corresponding RBM equation, plus relevant statistical analysis tools.
    \item Special high-level tools for propagation, e.g., shuffling propagators to cancel commutator errors and balancing auxiliary fields.
\end{itemize}

Future implementations include three-body forces, total angular momentum coupling, and coordinate dependence.

\bibliography{bib} 

\end{document}